\documentclass[useAMS,usenatbib,usegraphicx]{mn2e}

\usepackage{times,graphics,graphicx,amsmath,amssymb}
\usepackage[latin2]{inputenc}

\title[Detailed survey of the phase space around Nix and Hydra]
{Detailed survey of the phase space around Nix and Hydra}
\author[\'A. S\"uli, Zs. Zsigmond]{\'A. S\"uli and Zs. Zsigmond\\
Department of Astronomy, E\"otv\"os  University, Budapest, Hungary}
\date{Submitted 2009 May 25, Accepted 2009 June 18}

\pagerange{\pageref{firstpage}--\pageref{lastpage}} \pubyear{2009}

\begin{document}

\label{firstpage}

\maketitle

\begin{abstract}
We present a detailed survey of the dynamical structure of the phase space around the new moons of the Pluto--Charon system. The spatial elliptic restricted three-body problem was used as model and stability maps were created by chaos indicators. The orbital elements of the moons are in the stable domain both on the semimajor axis - eccentricity and - inclination spaces. The structures related to the 4:1 and 6:1 mean motion resonances are clearly visible on the maps. They do not contain the positions of the moons, confirming previous studies. We showed the possibility that Nix might be in the 4:1 resonance if its argument of pericenter or longitude of node falls in a certain range. The results strongly suggest that Hydra is not in the 6:1 resonance for arbitrary values of the argument of pericenter or longitude of node.
\end{abstract}


\section{Introduction}

In 1930 C. Tombaugh discovered Pluto, the ninth planet of the Solar system. From its discovery until 2006, Pluto was considered the Solar System's outmost planet. In the last two decades, however, many objects similar to Pluto were discovered in the outer solar system, notably the scattered disc object Eris, which is 27\% more massive than Pluto \citep{Brown2007}. On August 24, 2006 the IAU defined the term "planet" for the first time. This definition excluded Pluto, which the IAU reclassified as a member of the new category of dwarf planets along with Eris and Ceres.
\par
\begin{figure}
\includegraphics[width=1.0\linewidth]{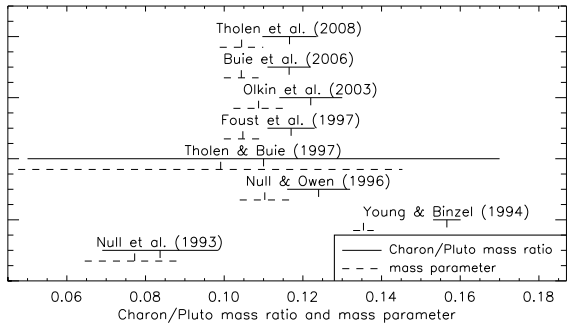}
\caption{Best-fit mass ratio $\hat{\mu}$ values for the Charon--Pluto system along with the corresponding mass parameters $\mu$. The horizontal bars span plus and minus one standard deviation from the best-fit values.}
\label{figure1}
\end{figure}

Pluto's first moon, Charon was found by \cite{Christy1978}, which greatly facilitated the study of Pluto, since that discovery made possible a more accurate determination of Pluto's mass. In Table \ref{table1} a chronology of mass ratios $\hat{\mu}=m_{\rm C}/m_{\rm P}$ and mass parameters $\mu=m_{\rm C}/(m_{\rm P}+m_{\rm C})$ is shown, where $m_{\rm P}$ and $m_{\rm C}$ are the masses of Pluto and Charon, respectively. Several authors have struggled to obtain this quantity from measurements of the barycentric wobble \citep{Null1993,Young1994,Null1996,Tholen1997,Foust1997,Olkin2003}. In addition the Pluto--Charon system is remarkable, since in the Solar system Charon is the largest moon relative to its primary, with the highest mass ratio of 0.1166 \citep{Tholen2008} (hereafter referenced to as T08). A visual representation of the mass ratios and the corresponding mass parameters is displayed in Figure \ref{figure1}. Horizontal solid lines give the error bars of the mass ratios with a vertical tick in the center denoting the best-fit value determined by the authors. Below each solid line the corresponding mass parameter is plotted as dashed lines, where the vertical tick denotes the mass parameter computed from the best-fit value of the mass ratio. We note that this is not at the center, as it is evident from the third column of Table \ref{table1}. The largest error bar is given for \cite{Tholen1997} since their observations were not optimized for the determination of the Charon/Pluto mass ratio. The mean of the mass ratios and mass parameters are given in the last row of Table \ref{table1}. In the works of \cite{Buie2006} (hereafter referenced to as B06) and T08 the masses were derived from two-body and four-body fits, respectively. From Table \ref{table1} one can see that the mass ratio has considerably evolved and reached by now a well established value. In the present work the dependence of the phase space on the mass parameter is studied.
\par
\begin{table}
\caption{Values for the Charon/Pluto mass ratios and mass parameters. In the first six rows values derived form the barycentric wobble are listed, while in the last two rows those calculated from orbital fits using the discovery of the small moons.}
\begin{center}
\begin{tabular}{lll}
\hline
Reference & $\hat{\mu} = \frac{M_{\rm C}}{M_{\rm P}}\times 10$ &
$\mu = \frac{\hat{\mu}}{1+\hat{\mu}} \times 10$ \\
\hline
\citet{Null1993}	& $0.837 \pm 1.47e-2$ & $0.7723^{+1.235e-2}_{-1.269e-2}$\\
\citet{Young1994}	& $1.566 \pm 3.50e-3$ & $1.3540^{+2.608e-3}_{-2.624e-3}$\\
\citet{Null1996}	& $1.240 \pm 8.00e-3$ & $1.1032^{+6.287e-3}_{-6.378e-3}$\\
\citet{Tholen1997}	& $1.100 \pm 6.00e-2$ & $0.9910^{+4.620e-2}_{-5.148e-2}$\\
\citet{Foust1997}	& $1.170 \pm 6.00e-3$ & $1.0474^{+4.783e-3}_{-4.835e-3}$\\
\citet{Olkin2003}	& $1.220 \pm 8.00e-3$ & $1.0873^{+6.310e-3}_{-6.400e-3}$\\
\hline
\citet{Buie2006}	& $1.165 \pm 5.50e-3$ & $1.0434^{+4.390e-3}_{-4.434e-3}$\\
\citet{Tholen2008}	& $1.166 \pm 6.90e-3$ & $1.0442^{+5.500e-3}_{-5.569e-3}$\\
\hline
Mean				& 1.1183              & 1.05537 \\
\hline
\hline
\end{tabular}
\label{table1}
\end{center}
\end{table}
\par
Subsequent searches for other moons around Pluto had been unsuccessful until mid May 2005, when two new moons were discovered \citep{Weaver2005,Weaver2006}. The discovery of Pluto's new moons, Nix (provisionally designated by S/2005 P2) and Hydra (provisionally designated by S/2005 P1) rendered the system even more interesting. They orbit the center of mass of the system, which is very close to the Pluto--Charon barycenter. The preliminary orbits computed by \cite{Weaver2005} (hereafter referenced to as W05) were based on just two observations separated by only three days, considerably less than a full orbit of either moon. Because of these constraints unique orbits could not be calculated from the available data, but the measured positions were consistent with nearly circular orbits in the orbital plane of Charon. On this assumption, preliminary orbital solutions yield $a=64700 \pm 850$ km and $P=38.2 \pm 0.8$ days for Hydra, and $a=49400 \pm 600$ km and $P=25.5 \pm 0.5$ days for Nix. Two-body orbit solutions for Nix and Hydra were computed by B06 using images taken of the Pluto system during 2002-2003 with the Hubble Space Telescope. Their use of data derived from prediscovery observations that span several orbits of all the moons made possible to compute unrestricted fits to the orbits of Nix and Hydra. According to the results, the orbital periods of Nix and Hydra are close to the ratio of 4:1 and 6:1 with that of Charon, respectively, indicating mean-motion resonance. In this paper two bodies are in mean-motion resonance when $n/n'=p/(p+q)$, where $n$ and $n'$ are their mean motions, $p$ and $q$ are small integers, where $q$ is the order of the resonance. The mean motions are measured in the sidereal system.
\par
Some of the major conclusions of B06 are as follows: ({\it i}) Nix and Hydra are in nearly circular orbits, with eccentricities of 0.0023 and 0.0052, respectively. ({\it ii}) The orbits of the small moons are almost coplanar with Charon's orbit. ({\it iii}) The orbital periods of Nix and Hydra are nearly commensurate with the period of Charon, but differ significantly from the exact ratios of 4:1 and 6:1, respectively. They argue that perhaps there are no resonances acting between the bodies. We note that in B06 the eccentricity of Charon was assumed to be zero; they argue that the eccentricity of the orbit of Hydra is significantly nonzero, unlike the orbits of Charon and Nix, which are consistent with zero eccentricity. As we will demonstrate the eccentricity of Charon substantially influences the phase space of the Pluto--Charon system.
\par
Although the two-body orbit solutions are good enough to provide satisfactory agreement with the 2002-2003 data, T08 studied whether the direct perturbations might be too strong to permit a sufficiently accurate extrapolation forward to the 2015 New Horizons spacecraft encounter with Pluto. Also, with adequate data, a formal four-body orbit solution should yield the mass for each member of the system.
\par
Their results of a four-body orbit solution for the Pluto system based on published observations have established 1$\sigma$ upper limits on the masses of Nix and Hydra. The masses and some other physical properties of the moons are shown in Table \ref{table2}. These limits place lower bounds on their geometric albedos and upper ones on their densities. The Charon/Pluto mass ratio was determined from the four-body orbit solution. The best-fit value of 0.1166 is almost identical to that of B06, since both team have used the same data.
\par
\begin{table}
\centering
\caption{Physical values of the moons, where $m$ is the mass, $\rho$ is the density, $D$ is the diameter and $\alpha$ is the visual geometric albedo (T08). Square brackets indicate assumed quantities.}
\begin{center}
\begin{tabular}{rlllll}
\hline
Name  &  $m$ [kg]             & $\rho$ & $D$ [km] & $\alpha$ (V) \\
\hline
Pluto & 1.304 $\times 10^{22}$ & 2.06   & [2294] & 0.61 \\
Charon& 1.520 $\times 10^{21}$ & 1.63   & 1212   & 0.34 \\
Nix   & 5.8 $\times 10^{17}$   & [1.63] & 88     & 0.08 \\
Hydra & 3.2 $\times 10^{17}$   & [1.63] & 72     & 0.18 \\
\hline
\hline
\end{tabular}
\end{center}
\label{table2}
\end{table}
\par
The work of T08 led to the conclusion that the orbits of Charon, Nix, and Hydra are not quite coplanar, and the latter moons' orbit planes precess around the system's invariable plane with periods of 5 and 15 years. The orbital eccentricities are nonzero but small for all three moons when measured in a barycentric reference frame. The results of T08 did not provide evidence on any mean motion resonances between the three moons confirming the study of B06.
\par
The eccentricity for Charon disagrees with the zero eccentricity published in B06. The erroneous value was the result of on the one hand using a software which was developed to fit highly eccentric orbit and on the other hand the use of the orbit published in \cite{Tholen1997}. It should be noted that the magnitude of Charon's eccentricity is very similar to the values determined from Pluto surface models of \cite{Tholen1997} (see Table III on page 251). In the case of the albedo model where only Pluto was considered, the computed eccentricity was $0.003 \pm 0.0005$ which agrees very well with the value of 0.0035 calculated in T08.
\par
To perform the four-body orbit solution all 22 parameters (18 orbital elements and 4 masses) were fitted simultaneously in the work by T08. During the computations, the solution settled into many different local minima, therefore without computing every possible orbit, it can not be guaranteed to find the absolute minimum. The currently available data are insufficient to produce a unique minimum. The possibility of stellar occultation opportunities and new Hubble Space Telescope observations will likely improve and refine the orbital elements of these new moons. Much more precise determination of the orbits and masses of Nix and Hydra will be possible as the New Horizons spacecraft approaches the Pluto system in 2015.
\par
Nagy, Süli \& Érdi (2006) studied the phase space of the Pluto--Charon system in the framework of the spatial circular restricted problem. The moons were treated as test particles and their semimajor axes, eccentricities and inclinations were varied. The computations were repeated for different initial mean anomalies. Their results showed that the region inside $\approx$ 42000 km is unstable, thus no moon could exist there. According to their results both moons reside in the stable region of the phase space of the Pluto--Charon system and the upper limit for the eccentricities of Hydra is 0.17, while it is 0.31 for Nix. In the semimajor axis - inclination plane the 4:1 and 6:1 resonances are clearly visible above $\approx 20^{\circ}$ and $\approx 35^{\circ}$, respectively. Unfortunately, these values could not serve to place a more stringent upper limit on these orbital elements.
\par
\begin{table*}
\centering
\begin{minipage}{160mm}
\caption[]{Orbital parameters derived by W05, Keplerian fits by B06 and those from four-body orbits solution by T08 typesetted in bold. The values are valid at epoch JD 2452600.5, mean equator and equinox of J2000. The parameters for the orbit of Charon are relative to Pluto while the orbits of Nix and Hydra are relative to the center of mass of the Pluto--Charon system. The values in parentheses in the semimajor axis column are given in units of the semimajor axis of Charon (A=1). Square brackets indicate assumed quantities.}
\center
\begin{tabular}{lrrrrrrrr}
\hline
Body		& $a$ [km] [A]&$e$     &$i$ [deg]& $\omega$ [deg]& $\Omega$ [deg] & $L$ [deg] & $M$ [deg] & $T$ [day] \\
\hline
Charon (B06)& 19571.4 (1)   	& 0.0    	 & 96.145		& --		& 223.046 		& 257.946		& 257.946		& 6.3872304  \\
Charon (T08)&{\bf 19570.3 (1)}  &{\bf 0.0035}&{\bf 96.168} 	&{\bf 157.9}&{\bf 223.054}	&{\bf 257.960}	&{\bf 237.006}	&{\bf 6.38720} \\
\hline
Nix (W05)	& 49400 (2.524) 	& [0.0]    	 & [96.145]		& --      	& --	 		& ?				& --			& 25.5    \\
Nix (B06)	& 48675 (2.487) 	& 0.0023 	 & 96.18		& 352.86  	& 223.14		& 123.14		& 267.14		& 24.8562 \\
Nix (T08)	&{\bf 49240 (2.516)}&{\bf 0.0119}&{\bf 96.190}	&{\bf 244.3}&{\bf 223.202}	&{\bf 122.7}	&{\bf 15.198}	&{\bf 25.49} \\
\hline
Hydra (W05)	& 64770 (3.309)		& [0.0]    	 & [96.145]		& --		& --			& ?				& --			& 38.2    \\
Hydra (B06)	& 64780 (3.310)		& 0.0052 	 & 96.36		& 336.927	& 223.173		& 322.71		& 122.61		& 38.2065 \\
Hydra (T08)	&{\bf 65210 (3.332)}&{\bf 0.0078}&{\bf 96.362}	&{\bf 45.4}	&{\bf 223.077}	&{\bf 322.4}	&{\bf 53.923}	&{\bf 38.85} \\
\hline
\hline
\end{tabular}
\label{table3}
\end{minipage}
\end{table*}
\par
The main goal of this paper is to extend the previous investigations of \cite{Nagy2006} using the spatial elliptic restricted problem. In their work the eccentricity of Charon was zero and the mass parameter of the system was 0.130137. Since then both of these values have been updated and also the orbital elements of Nix and Hydra have been refined as it can be seen from Table \ref{table3}. The relevant changes are the following: ({\it i}) the semimajor axes of both moons have been changed, ({\it ii}) Nix's eccentricity is significantly different from zero, and Hydra's eccentricity has increased by a factor of 150\% ({\it iii}) the argument of periapses have completely different values. The most important change is the transition from the circular to the elliptic case. In the circular restricted problem there exists a first integral of motion, i.e. the Jacobi-integral, which does not exist when introducing the eccentricity. Due to the disappearance of the Jacobi integral the phase space structure changes qualitatively.
\par
With that said, the orbital elements tabulated in Table \ref{table3} are subject to change as new observations become available. Using stability maps computed in advance for a large set of orbital parameters has the advantage that the stability properties of the moons can be easily established when the orbital parameters are modified. Moreover the error bars can be readily combined with the stability maps giving extra information about the motion.
\par
In Section 2 we describe the investigated model and give the initial conditions used in the integrations. The applied numerical methods are briefly explained in Section 3. The results are presented in Section 4. Section 5 is devoted to the conclusions.

\section{Model and initial conditions}

Since the orbital radii of the moons are much smaller than the Hill radius ($\approx 8.0 \times 10^6$ km) of the Pluto--Charon system, the moons are deep in Pluto's gravitational well, the perturbations by the Sun can be ignored, as did \cite{Lee2006} and T08.
\par
To study the structure of the phase space of the Pluto--Charon system we applied the model of the spatial elliptic restricted three-body problem. We integrated the dimensionless equations of motion. An obvious advantage of using such equations is that the results are independent of the exact value of the semimajor axis of Charon. The unit of length was chosen such that the separation of Pluto and Charon (the primaries) is unity, i.e. the semimajor axis of Charon $a_1=1$ A in all computations. Let the unit of mass be the sum of the primaries, i.e. $m_{\rm P}+m_{\rm C}=1$, where $m_{\rm P}$ and $m_{\rm C}$ are the masses of Pluto and Charon, respectively. Let the unit of time be chosen such that $k^2(m_{\rm P}+m_{\rm C})=1$, where $k$ is the gravitational constant. The orbital plane of the primaries was used as reference plane, in which the line connecting the primaries at $t=0$ defines a reference $x$-axis. The longitude of the ascending node $\Omega$ is the angle between the line of nodes and the $x$ axis. The argument of pericenter $\omega$ is measured between the line of nodes and the radius vector of the pericenter. The values of $\Omega=0$, $\omega=0$ and the mean anomaly $M=0$ places the test particle on the $x$ axis at a distance of $a(1-e)$ from the barycenter, where $a$ is the semimajor axis and $e$ is the eccentricity of the test particle.
\par
The initial orbital elements of Charon are given in Table \ref{table3}, where those of T08 were used. Hereafter the orbital elements of Charon will be denoted by subscript 1, i.e. $e_1$ is the notation for Charon's eccentricity.
\par
Though Nix and Hydra are almost coplanar with Charon, still we study the problem more generally by considering the effect of non-zero inclinations on the orbital stability. The mass parameter $\mu=0.104424$ was chosen according to the value of 0.1166 published in T08 (see Table \ref{table1}).
\par
To examine the phase space in the vicinity of Nix and Hydra separately we varied the initial orbital elements of the test particles. Stability maps were created for the $(a-e)$ and $(a-i)$ orbital element space for both moons for different eccentricity $e$, inclination $i$, argument of pericenter $\omega$ and longitude of node $\Omega$. The values of the mean anomalies $M$ were kept constant in all the simulations and are given in Table \ref{table3}. To compute the $(a-e)$ and $(a-i)$ stability maps the semimajor axis was varied with a stepsize of $10^{-3}$ in the intervals which are given in Table \ref{table4}. For each computed $(a-i)$ map the inclination was changed from 0 to 45$^{\circ}$ with $\Delta i=1^{\circ}$ and for each $(a-e)$ map the eccentricity varied from 0 to 0.4 with $\Delta e=10^{-3}$.
\par
The $(a-e)$ stability maps were computed for different parameters of the moons. For each $([i]_k,[\omega]_l)$ and $([i]_k,[\Omega]_l)$ pair an $(a-e)$ stability map was computed, where
\begin{eqnarray}
\left[ i \right]_k&=& k\Delta i',\quad k=0, \ldots,4,\nonumber \\
\left[ \omega \right]_l&=&22.^{\circ}9+l\Delta \omega',\quad [\Omega]_l=l\Delta \Omega',\quad l=0,\ldots,7,\nonumber
\end{eqnarray}
where $\Delta i'=10^{\circ}$\,, $\Delta \omega'=\Delta \Omega' = 45^{\circ}$. Here we used the [] notion to distinct the variables from those of Charon. Computation of each combination of the $(k,l)$ integer pair results in $2 \times 40$ stability maps per moons.
\par
Similarly the $(a-i)$ stability maps were computed for each $([e]_k,[\omega]_l)$ and $([e]_k,[\Omega]_l)$ pair, where
\begin{equation}
\left[ e \right]_k= k\Delta e',\quad k=0, \ldots,4.\\
\end{equation}
where $\Delta e'=0.1$. The details are summarized in Table \ref{table4}.
\par
In total more than 14 million orbits were calculated, which resulted in $2\times80=160$ stability maps for each moon. Due to the very small stepsize in $a$ and $e$, each $(a-e)$ stability map corresponds to more than $6 \times 10^{4}$ initial conditions, thus providing a very fine resolution.
\par
\begin{table}
\centering
\caption{The initial orbital elements for the test particles. In the upper part the intervals of the semimajor axes along with the stepsizes $\Delta$ are listed for the stability maps $(a-e)$ and $(a-i)$. The lower part shows the intervals (I) for the orbital elements $[e]_k\,,[\omega]_l$ and $[\Omega]_l$ and the respective stepsizes ($\Delta'$).}
\begin{center}
\begin{tabular}{lrrrr}
\hline
Map		& $a$ [A]     &  $e$      & $i$ [deg]&\\
\hline
Nix		& [2.40,2.64] & [0,0.4]   & [0,45]	&\\
Hydra	& [3.22,3.38] & [0,0.4]   & [0,45]	&\\
$\Delta$&  $10^{-3}$  & $10^{-3}$ &      1	&\\
\hline
		& $[e]_k$         & $[i]_k$ [deg] & $[\omega]_l$ [deg]& $[\Omega]_l$ [deg]\\
\hline
I		& [0,0.4]	  & [0,40]	  & [22.9;337.9]  & [0;315]\\
$\Delta'$&    0.1	  &    10     &        45     &     45 \\
\hline
\hline
\end{tabular}
\end{center}
\label{table4}
\end{table}
\par
The above orbital elements refer to a barycentric reference frame, where the mass of the barycenter is $m_{\rm P}+m_{\rm C}$. By the usual procedure we calculated the barycentric coordinates and velocities of the test particle and then transformed them to a reference frame with Pluto in the origin. In the numerical integrations we used the latter coordinates and velocities.
\par
For the integration of the system we applied an efficient variable-timestep algorithm, known as Bulirsch--Stoer integration method. The most important feature of this algorithm for simulations is that it is capable of keeping an upper bound on the local errors introduced due to taking finite timesteps by adaptively reducing the step size when interactions between the particles increase in strength. The parameter $\epsilon$ which controls the accuracy of the integration was set to $\epsilon=10^{-10}$. The orbits were integrated for $10^4$ Charon's period (hereafter $T_{\rm C}$).

\section{Methods}

To compute the stability maps, the method of the maximum eccentricity (ME), the Lyapunov characteristic indicator (LCI) and the relative Lyapunov indicator (RLI) were used as tools for stability investigations of the massless bodies representing the small moons.
\par
The ME method uses as an indication of stability a straightforward check based on the eccentricity. This action-like variable shows the probability of orbital crossing and close encounter of two bodies and therefore its value provides information on the stability of orbits. This simple check was already used in several stability investigations, and was found to be a powerful indicator of the stability character of orbits \citep{Dvorak2003,Suli2005,Nagy2006}. In this work we define ME as follows
\[
\mathrm{ME} = \mathop{\mathrm{max}}\limits_{t\in[0,10^4]T_{\rm C}}(e).
\]
\par
As a complementary tool, we computed also the LCI, a well-known chaos indicator. The LCI is the finite time approximation of the largest Lyapunov exponent, which is described in detail in e.g. \cite{Froeschle1984}. The definition of the LCI is given by
\[
\mathrm{LCI}(t,x_0,\xi_0) = \frac{1}{t} \log \Vert \xi(t) \Vert, \nonumber
\]
where $x_0$ is the initial condition of the orbit and $\xi(t)$ is the solution of the first order variational equations. The function $\mathrm{LCI}(t,x_0,\xi_0)$ measures the mean rate of divergence of the orbits.
\par
In \cite{Sandor2000,Sandor2004} the difference between the LCIs of two neighbouring orbits was introduced:
\[
\Delta L(x_0,\xi_0,\Delta) = \vert \mathrm{LCI}(x_0,\xi_0,t) - \mathrm{LCI}(x_0+\Delta, \xi_0,t) \vert,\nonumber
\]
where $\Delta$ is the distance between two close orbits. This quantity measures the fluctuations of the curve of the LCI. To smooth the time evolution of $\Delta L$ its time average is computed, which is the definition of the RLI:
\[
\mathrm{RLI}(x_0,\xi_0,\Delta) =\frac{1}{t} \sum \vert \Delta L(x_0,\xi_0,\Delta) \vert,\nonumber
\]
The definition contains $\Delta$ as free parameter, which must be chosen small enough to reflect the local properties of the flow in the phase space. It was shown, (\cite{Sandor2000,Sandor2004}) that the choice of this parameter in a quite large interval ($\Delta \in [10^{-14},10^{-7}]$) does not modify essentially the behaviour of the RLI. This method is extremely fast to determine the ordered or chaotic nature of orbits and the method is sensitive for showing the resonant structure on stability maps \citep{Sandor2006}. In our computations $\Delta=10^{-10}$ was used.
\par
The three methods are not equivalent, however they complete each other. For example, the ME of the Earth is small, indicating stability, although we know from numerical experiments that in fact the Earth is moving on a chaotic trajectory with a small but nonzero Lyapunov exponent. Therefore the ME detects macroscopic instability (which may even result in an escape from the system), whereas the RLI and the LCI are capable to indicate microscopic instability.
\par
According to notation. (\ref{eq1}) the stability map is prepared in such a way, that for each set of the described initial conditions the LCI, RLI and ME value of the corresponding orbit of the test particle is computed and plotted in the $(a-e)$ or $(a-i)$ parameter plane.

\section{Results}

In this Section the $\mathrm{ME}_-$ stability maps are presented (Figs \ref{figure2}--\ref{figure7}), where $\mathrm{ME}_-$ is defined as follows:
\[
\mathrm{ME}_- = \left\{\begin{array}{ll}
1  & \text{if ME = 1} \\
\mathrm{ME} - e_0 & \text{otherwise}\\
\end{array}\right.,
\]
where $e_0$ is the initial eccentricity of the test particle. We used the above definition in order to display the maximum change in $e$ in the course of the integration and to be able to compare orbits with different $e_0$. For each $(a,e)$ or $(a,i)$ point the $\mathrm{ME}_-$ is computed and depending on its value a greyscale dot is plotted. Darker shades correspond to very low values of the $\mathrm{ME}_-$ and regular behaviour of the test particle, while lighter shades indicate larger $\mathrm{ME}_-$. In order to better visualize the structures in the stability maps the scale for the $(a-e)$ figures extends from zero to 0.2 for Nix and 0.1 for Hydra, for the $(a-i)$ maps to 0.2 and 0.05, respectively. In the $(a-e)$ maps the thick black solid curve in the white upper region is a contour line where ME is approximately equals to 0.9 and marks the upper boundary of stable region. Above the curve ME is practically 1.0 while below it ME rapidly decreases. The thin dashed contour line right below the thick one corresponds to ME=0.4 and in all regions below this curve the $\mathrm{ME}_-$ is less than 0.2 and 0.1 for Nix and Hydra, respectively. In the $(a-i)$ maps ME is less than 0.4 in all cases.
\par
On the figures white solid curves are contour lines: along these curves the $\mathrm{ME}_-$ has a constant value. These values were chosen in such a way that the curves draw the approximate boundary of the most prominent structure and to make comparison possible. The white numbers at the upper end of these curves are the corresponding $\omega$ or $\Omega$ values. \cite{Murray99} derived the maximum libration zone of resonances as a function of the semimajor axis and eccentricity using an analytical model based on the circular restricted three-body problem. These zones for $q \ge 2$ resonances have a V-shape on the $(a-e)$ plane. The V-shapes are well represented by the white contour curves. Using a different, but reasonable $\mathrm{ME}_-$ value would provide a similar V-shape. The chosen values to plot the contour curves are based on several tests.
\par
The stability maps based on the LCI and RLI values are essentially the same as those based on $\mathrm{ME}_-$, therefore these maps are not presented.

\subsection{The mass parameter}

The mass parameter used by \citet{Nagy2006} differs by 25\% from the present best-fit value therefore we studied the system with different $\mu$ values. For the sake of comparison four runs were performed, one for $\mu=0.130137$ and one for $\mu=0.104424$ in the $(a-e)$ phase space in the vicinity of both Hydra and Nix. The stability map in the vicinity of Nix for $\mu=0.104424$ and $e_1=0$ is shown in the upper panel of Figure \ref{figure2}. The phase space structures for the two $\mu$ parameters (the one for $\mu=0.130137$ is not shown) are almost identical, they are only shifted along the horizontal axis: structures for the lower mass parameter are closer to Pluto. This is a consequence of the changes in the coordinates of the barycenter. This shift amounts approximately to 0.005 A. The ME=0.9 contour line is almost the same although in the boundary zone between the unstable and stable region the maximum difference in $\mathrm{ME}_-$ computed using the two mass parameters can reach values as high as $\approx 1$. This large difference is a natural consequence of the chaotic nature of orbits emanating from this region. Using longer integration time would cease these large differences. The conclusion is that in this mass parameter region of $\approx 0.1$, 25\% change in $\mu$ shifts the phase space structures along the horizontal axis. This shift could be important in the close vicinity of mean motion resonances, like in the case of Hydra where the lower end of the 6:1 resonance for $\mu=0.130137$ is very close to T08 but it shifts away for $\mu=0.104424$ (see Fig. \ref{figure3}).

\subsection{Charon's eccentricity}

Since the Jacobi-integral does not exist in the elliptic problem therefore the phase space structure is qualitatively distinct in the two models. In order to visualize the effect of the disappearance of the Jacobi-integral, the circular and the elliptic cases for Nix are plotted in Fig. \ref{figure2} and for Hydra in Fig. \ref {figure3} where the circular case is shown in the upper the elliptic case in the lower panel. For Nix two major differences comparing with the circular case are the following: ({\it i}) the unstable zone is much larger in the elliptic case beginning from $e \approx 0.17$ onward and ({\it ii}) the center of the 4:1 resonance is shifted from $a \approx 2.544$ A to $a \approx 2.564$ A. The shape of the resonance did not change significantly but the fine inner texture is distinct. For Hydra the upper unstable zone extends too, the center of the resonance moves closer to Pluto and its shape changes completely. In both cases the resonance becomes stronger for low eccentricities in the elliptic case.

\subsection{The $(a-e)$ stability maps}

\begin{figure}
\includegraphics[width=1.0\linewidth]{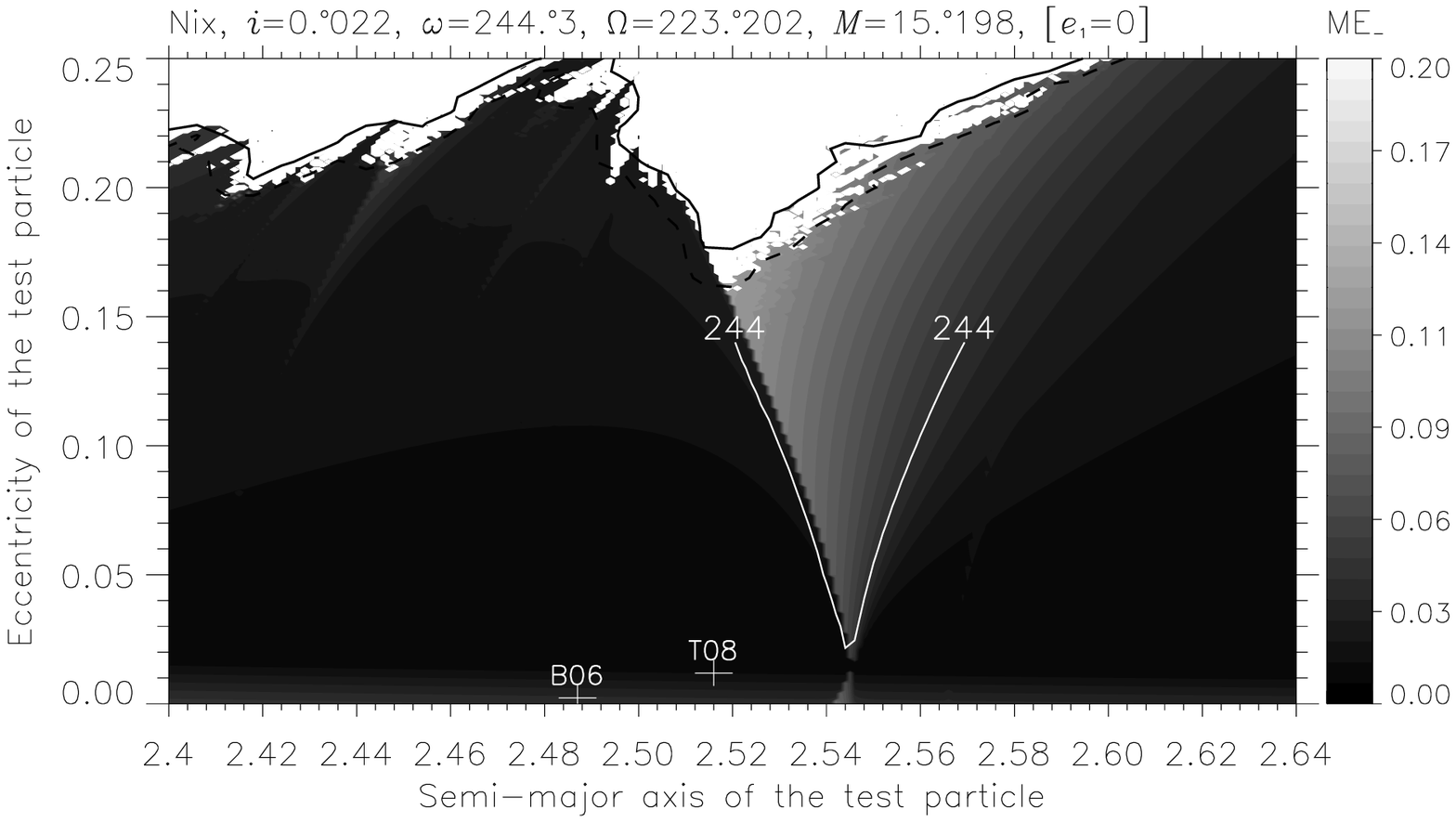}
\includegraphics[width=1.0\linewidth]{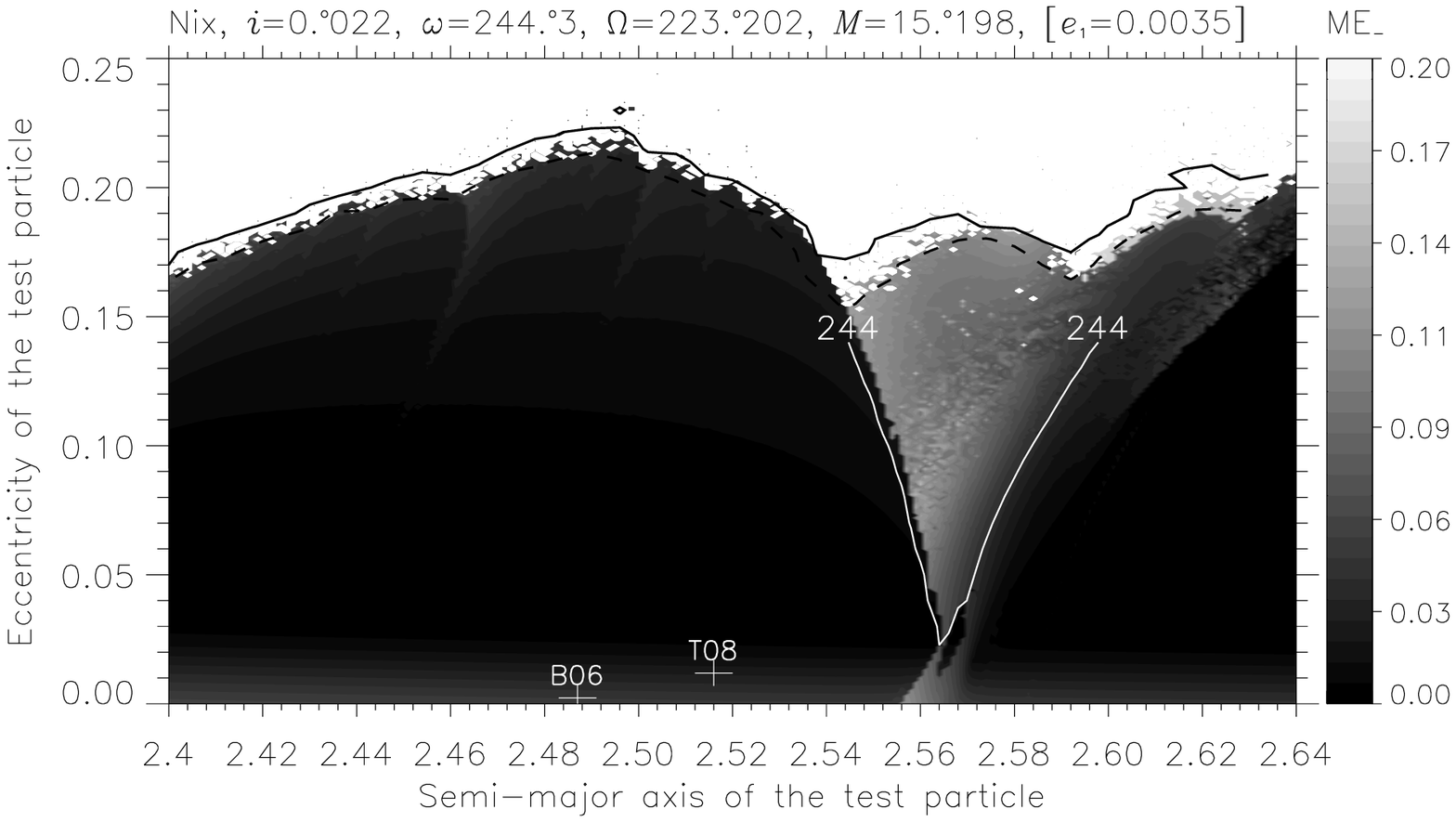}
\caption{The $(a-e)$ stability map for Nix with $\mu=0.104424$. In the upper panel the circular ($e_1=0$), in the lower panel the elliptic case ($e_1=0.0035$) is displayed. Orbits emanating above the upper solid thick black curve are unstable orbits with ME$\ge 0.9$. The plus signs correspond to the orbital elements published by B06 and T08. Along the contour curves $\mathrm{ME}_-=0.04$.}
\label{figure2}
\end{figure}

\begin{figure}
\includegraphics[width=1.0\linewidth]{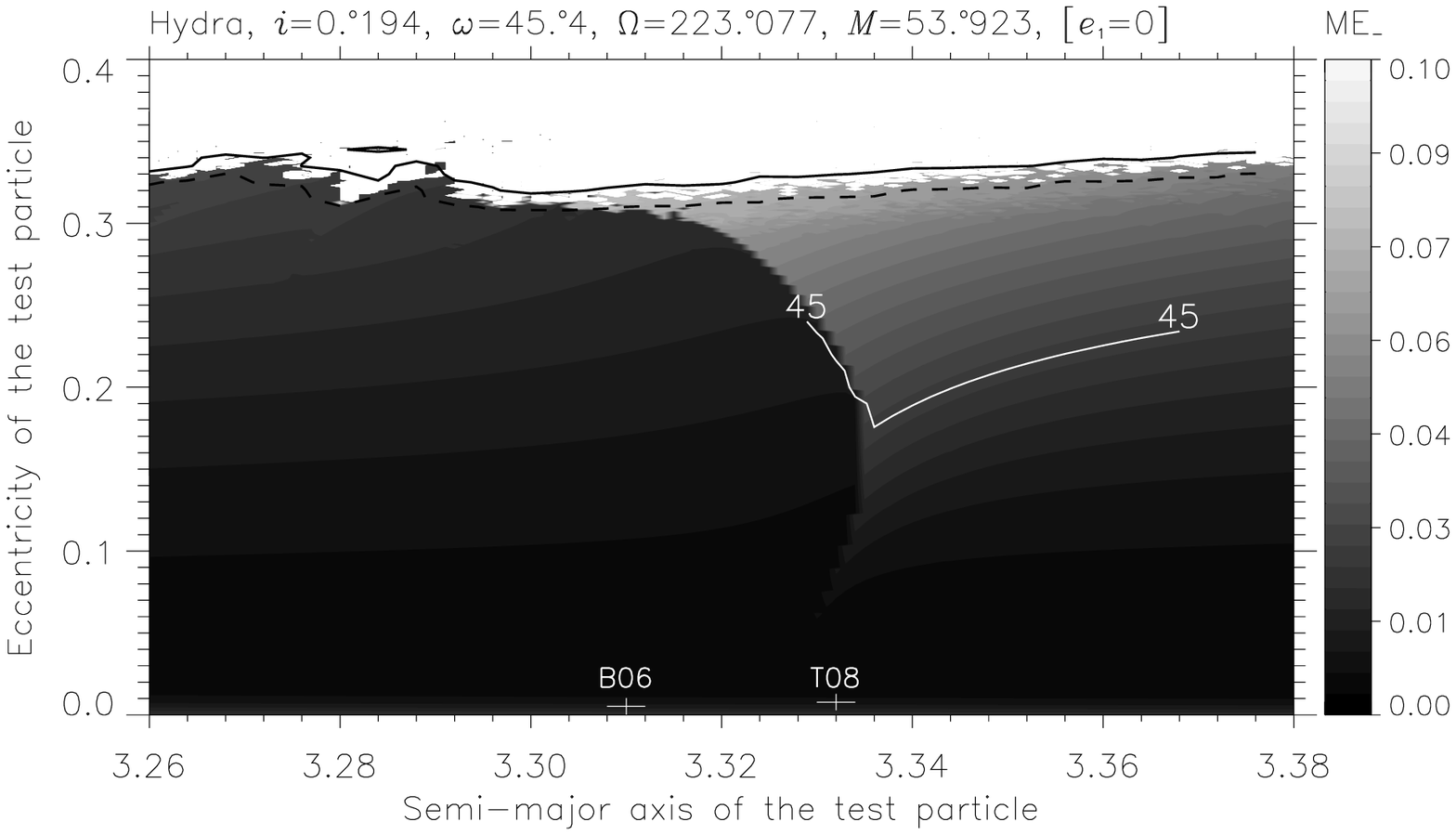}
\includegraphics[width=1.0\linewidth]{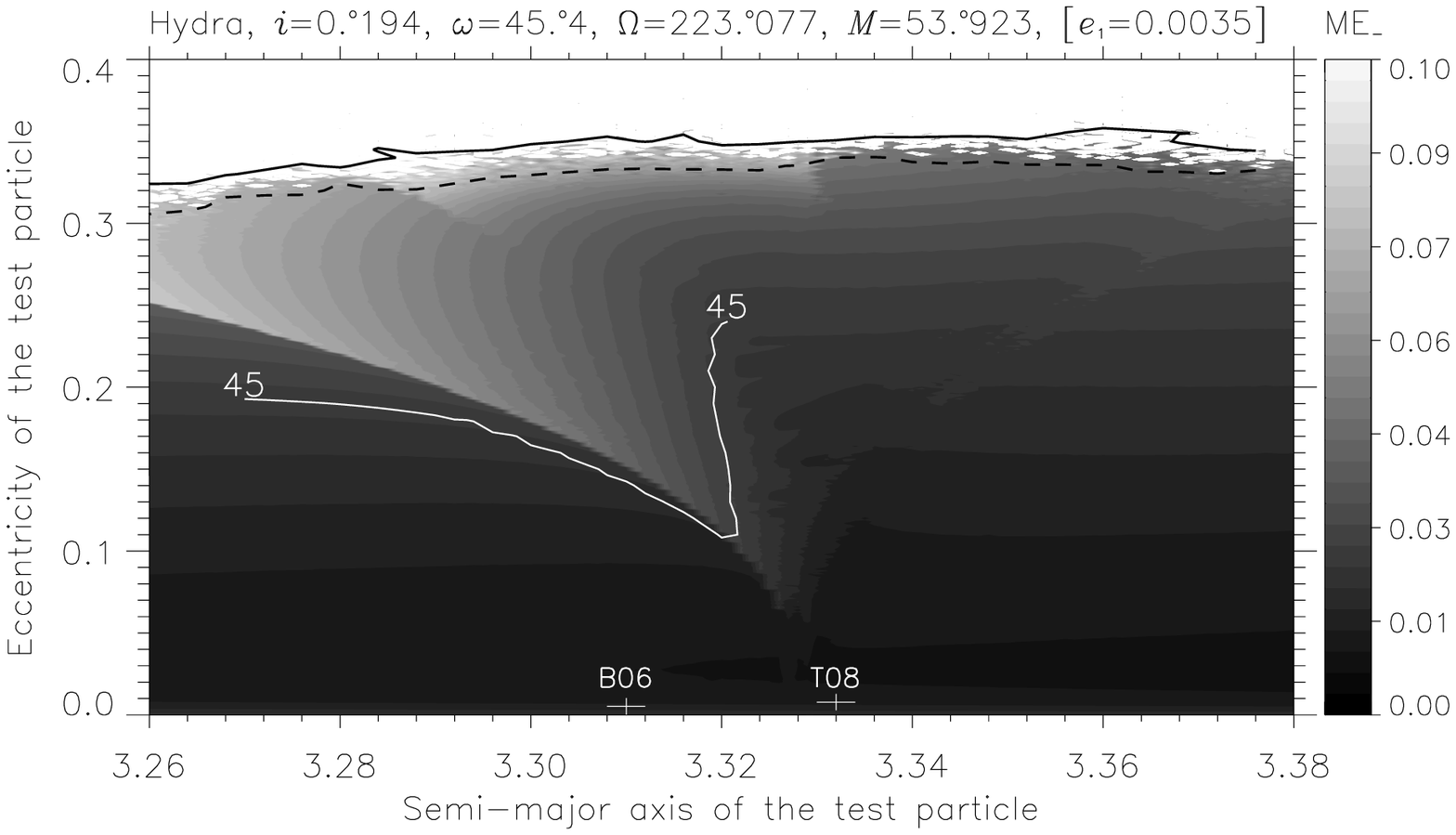}
\caption{The $(a-e)$ stability map for Hydra with $\mu=0.104424$. The V-shape structure corresponds to the 6:1 resonance. Along the contour curves $\mathrm{ME}_-=0.025$. See explanation of Fig. \ref{figure2}.}
\label{figure3}
\end{figure}

The $(a-e)$ stability maps are shown in the lower panels of Fig. \ref{figure2} for Nix and in Fig. \ref{figure3} for Hydra. The orbital elements of the moons were taken from Table \ref{table3} row T08. Each of the two figures is dominated by a V-shaped gray structure, corresponding to the 4:1 and 6:1 mean motion resonances between Charon and Nix and Hydra, respectively. These resonances can represent either ordered (stable for infinite time), or weakly chaotic (which may become unstable after very long time) behaviour. The location of the plus signs corresponds to the orbital elements published by B06 and T08. The size of the plus sign is not proportional to the error bars of the $a$ and $e$ values. The white curve is a contour line corresponding to $\mathrm{ME}_-$ value of 0.04 for Nix and 0.025 for Hydra.
\par
From Fig. \ref{figure2} it is clearly visible that the present position of Nix is not in the 4:1 mean motion resonance, although it is closer than the position of B06. Our results are in agreement with B06 and T08 since none of these works have identified any resonant arguments between Nix and Charon. We note that there is a sharp jump in the $\mathrm{ME}_-$ values along the left separatrix while the $\mathrm{ME}_-$ decreases smoothly as a function of $a$ as the right separatrix is approached.
\par
Fig. \ref{figure3} shows the stability map for Hydra. The V-shape 6:1 resonance is clearly visible and its lower peak is very close to the present position of Hydra. However, the $\mathrm{ME}_-$ values are very small, the white contour curve indicates the region where $\mathrm{ME}_-$ is above 0.025. The $\mathrm{ME}_-$ in the close vicinity of Hydra is less than 0.02 indicating that this high order resonance is very weak in the close proximity of Hydra. Again this result confirms those of B06 and T08.
\par
\begin{figure*}
\includegraphics[width=1.0\linewidth]{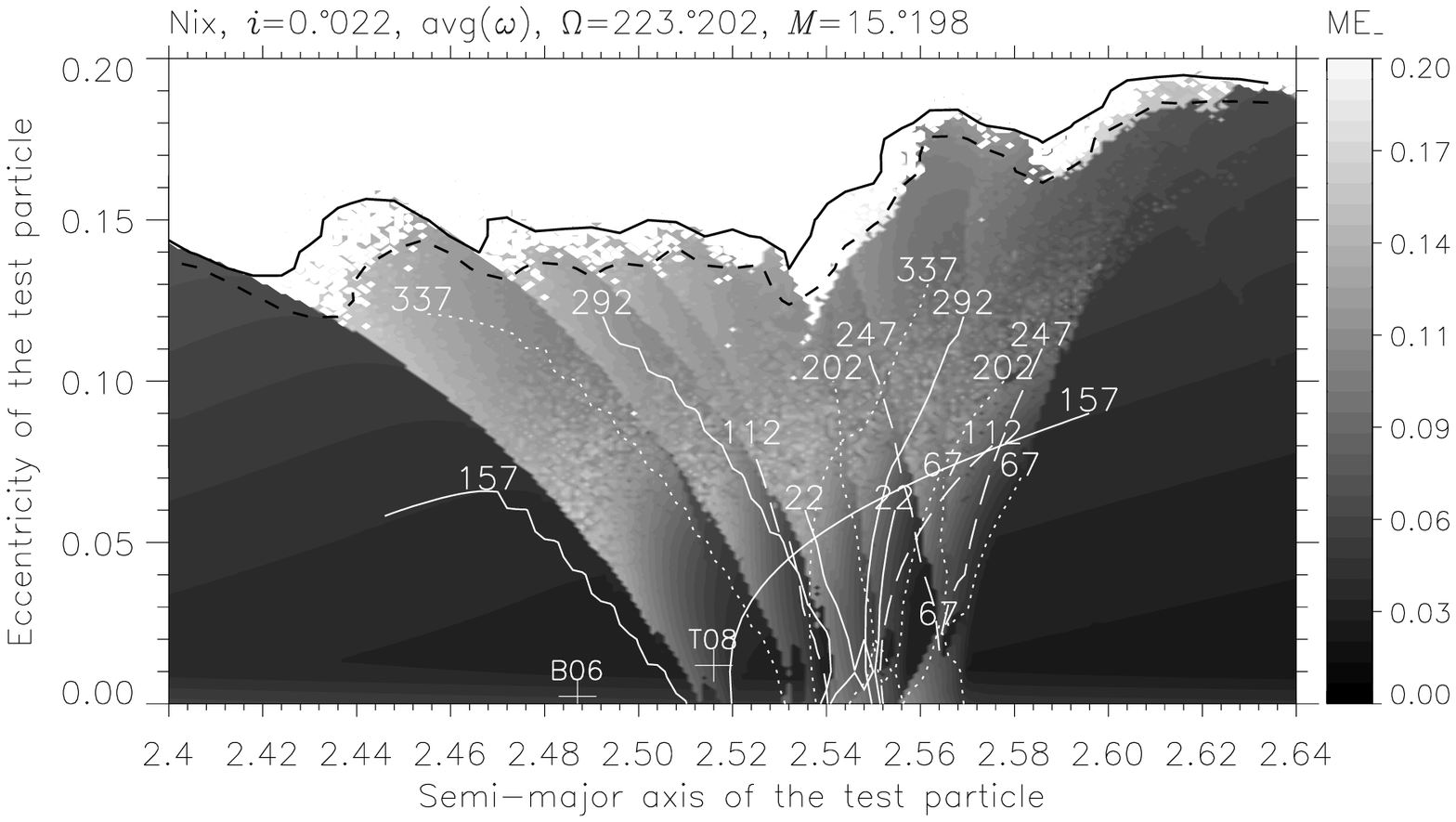}
\includegraphics[width=1.0\linewidth]{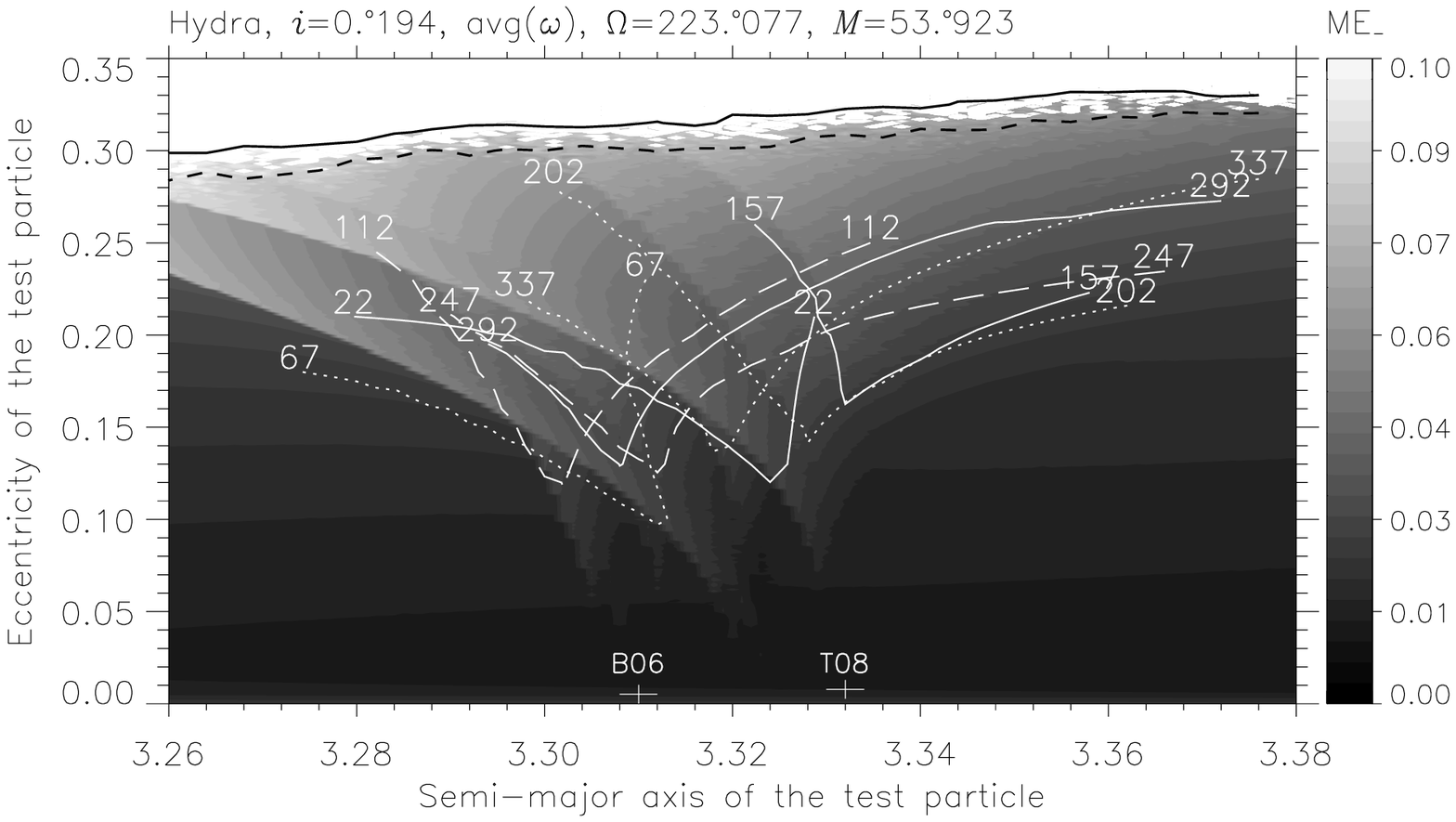}
\caption{The $(a-e)$ stability map for 8 $\omega$ values for Nix (upper) and Hydra (lower). The white curves correspond to contour lines indicating the shape of the resonances as they would appear on individual maps for the $\omega$ displayed at the upper end of the curves. In the upper panel along the contour curves $\mathrm{ME}_-=0.04$. In the case of Hydra along the contour curves $\mathrm{ME}_-=0.025$.}
\label{figure4}
\end{figure*}

In the next step we investigated the orbits systematically by changing the initial orbital elements of the test particle as described in Section 3 (see also Table \ref{table4}). In Fig. \ref{figure4} the results are summarized for the 8 values of $\omega$. These "maximum" stability maps were obtained as follows: for each $(a,e)$ point we plotted the maximum selected from the $\mathrm{ME}_-$ values computed for the 8 different $\omega$. The white curves on the figures approximately denote the location of the resonances as they appear on individual stability maps, like in Figs \ref{figure2} and \ref{figure3}.
\par
In Fig. \ref{figure4} the upper panel shows the parameter space around Nix where the contour lines belong to $\mathrm{ME}_- = 0.04$. In the case of $\omega=157.^{\circ}9$ Nix is inside the 4:1 resonance, and between the curves the $\mathrm{ME}_-$ values are significantly higher than for the other values of $\omega$. In this case the pericentrum of Nix is the same as that of Charon. The detailed study of this configuration will be carried out in a later work. The former B06 position is obviously outside of the resonance confirming the findings of B06. The location, size and shape of the resonance strongly depends on the $\omega$ of Nix. From Fig. \ref{figure4} it is evident that the effect of the 4:1 resonance is not negligible for even very small eccentricities.
\par
In Fig. \ref{figure4} the lower panel displays the parameter space around Hydra where the contour lines belong to $\mathrm{ME}_- = 0.025$. The ME=0.9 curve depends approximately linearly on $a$, starting from $e=0.3$ at $a=3.26$ A and ending at $e=0.33$ at $a=3.38$ A. The 6:1 mean motion resonance is visible for all $\omega$, but it can not be traced down below $e \approx 0.05$ for any value of the pericenter. The resonance is the strongest for $\omega = 67^{\circ}$ in the sense that its effect is visible for the lowest $e \approx 0.05$. The present position of Hydra is closest to the resonance when $\omega=22^{\circ}$: if the upper limit of the scale is reduced to 0.03 than the resonance lowest peak is just above the T08 plus sign. As it is obvious from Fig. \ref{figure4} the 6:1 mean motion resonance is very weak in the vicinity of the moon, no sign of it can be observed around T08 or B06.
\par
The $\omega$ published in T08 differs by $108^{\circ}$ from that of B06 in the case of Nix, and by $68 ^{\circ}$ in the case of Hydra (see Table \ref{table3}). Since the eccentricity is small for all three moons the value of $\omega$ is hard to establish accurately. New and more accurate orbital solution based on data yielded by future observations of the system could result in such $\omega$ that would place Nix in the resonance. In the case of Hydra this is questionable since the 6:1 resonance is very weak for low eccentricities. The same applies to the nodes because the moons are nearly coplanar, however the value of $\Omega$ changed only a few arcminutes.
\par
We note that the switch from 2D to 3D in the case of small inclination $(i \le 0.^{\circ}1)$ does practically not induce any observable changes in the $(a-e)$ parameter space. Computations were performed for zero inclination and the comparison of the stability maps have revealed only tiny differences.
\par
We briefly give a summary for higher inclination in the following. Nix's maximum stability map for $i=10^{\circ}$ indicates stronger effects of the 4:1 resonance and the expansion of the unstable zone which starts already from $e=0.1$. The resonances are wider and closer to Pluto and to the T08 position, which is now inside more resonance structures indicating the potential of active resonance. Increasing the $i$ up to $20^{\circ}$ further shrinks the stable domain which ends at $e\approx 0.06$. The resonance moves even closer to Pluto and gain in power. The maximum stability maps for $i=30^{\circ}$ and $40^{\circ}$ show that the lower border of the unstable zone stays around 0.06 and the resonance structures move closer to the barycenter.
\par
Hydra's maximum stability map for $i=10^{\circ}$ is similar to Fig. \ref{figure4}: the ME=0.9 contour curve is practically the same as well as the resonant structures which have become slightly more prominent. For $i \ge 20^{\circ}$ the resonant structures can be followed down to the $a$-axis and like in the case of Nix they gain in power and shift closer to Pluto: in the $i=40^{\circ}$ case the T08 position is outside the resonance. The upper limit of the stable zone is $e\approx 0.28$ at $i=40^{\circ}$: large increase in the inclination did not significantly reduce the size of the stable domain around the moon.
\par
For both moons the "maximum" stability maps for the 8 values of $\Omega$ were obtained on the analogy of Fig. \ref{figure4}. For $i=i_{\rm N}$ and $i=i_{\rm H}$ the results agree quite well with the ones obtained by $\omega$: each structure on Fig. \ref{figure4} has a counterpart with $\Omega_{\rm N} \approx \omega_{\rm N}-22^{\circ}$, and $\Omega_{\rm H} \approx \omega_{\rm H}-180^{\circ}$. The explanation of these simple relationships is the following: in the limit as $i \rightarrow 0$ the orbital plane coincides with the reference plane and we have $\varpi = \omega+\Omega$, where $\varpi$ is the longitude of the pericenter. In the case of Nix from the equation
\[
\varpi_{\omega}=\varpi_{\Omega},
\]
where $\varpi_{\omega}=\omega_i+\Omega_{\rm N}$ and $\varpi_{\Omega}=\omega_{\rm N}+\Omega_i$ it follows that \mbox{$\Omega_i=\omega_i-21^{\circ}$}, where $\rm{i}=1,\ldots,8$. The equation yields for Hydra $\Omega_i=\omega_i-182^{\circ}$.
\par
As a consequence the stability map belonging to \mbox{$\Omega=135^{\circ}$} indicates that Nix might be in the 4:1 resonance. In the case of nearly coplanar orbits even small details are similar on both figures. The above relationships are also true for $i=10^{\circ}$ for the dominant features but the details are already distinct. The results disagree for higher inclination.

\subsection{The $(a-i)$ stability maps}

\begin{figure}
\includegraphics[width=1.0\linewidth]{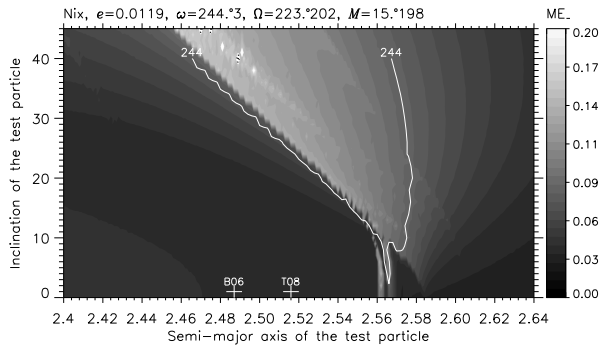}
\caption{The $(a-i)$ stability map for Nix with initial conditions taken from T08. Along the contour curves $\mathrm{ME}_-=0.08$.}
\label{figure5}
\end{figure}

\begin{figure}
\includegraphics[width=1.0\linewidth]{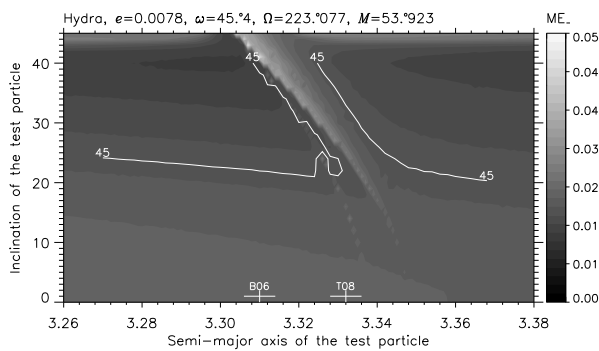}
\caption{The $(a-i)$ stability map for Hydra with initial conditions taken from T08. Along the contour curves $\mathrm{ME}_-=0.017$.}
\label{figure6}
\end{figure}

The $(a-i)$ stability maps are shown in Fig. \ref{figure5} for Nix and in Fig. \ref{figure6} for Hydra. The orbital elements of the moons were taken from Table \ref{table3} row T08. In the case of Nix the contour line belongs to $\mathrm{ME}_- = 0.08$; we note that this value is the double than those belonging to the $(a-e)$ maps. In accordance with Fig. \ref{figure2} the results displayed in Fig. \ref{figure5} confirms that the present position of Nix is not in the 4:1 mean motion resonance, but in a region where $\mathrm{ME}_-$ is very small indicating stable motion. The remarkable feature of the resonance is a vertical, approximately 0.008 A wide and 10 degree high column at $a=2.564$ A. Let us note that the shape of the resonance is reminiscent of a bee's abdomen with a sting. As before there is a sharp jump in the $\mathrm{ME}_-$ values along the left side and $\mathrm{ME}_-$ decreases smoothly towards the right separatrix.
\par
For Hydra the top of the scale of $\mathrm{ME}_-$ is 0.05 and the white contour line belongs to $\mathrm{ME}_- = 0.017$. The 6:1 mean motion resonance structure is observable only for $i \ge 20^{\circ}$ where the maximum of $\mathrm{ME}_-$ is less than 0.05 indicating that the resonance is very weak. In accordance with Fig. \ref{figure3} Hydra is not in the 6:1 mean motion resonance. The two regions enclosed by the white curves on the right and left side of the resonance contain values that are less than 0.017. It is interesting that two local minima appear at $i \approx 40^{\circ}$ on both sides of the resonant structure.
\par
\begin{figure*}
\includegraphics[width=1.0\linewidth]{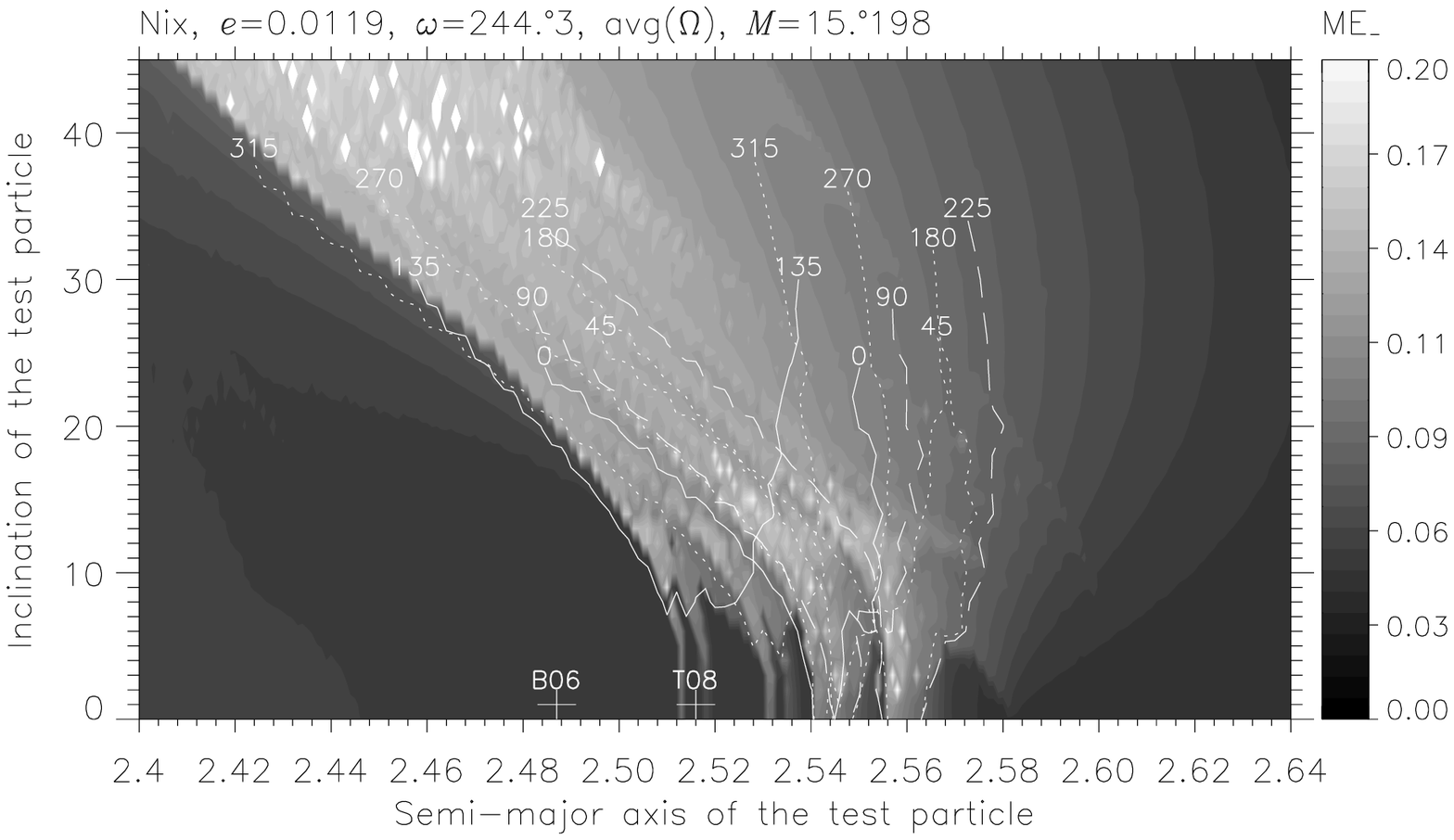}
\includegraphics[width=1.0\linewidth]{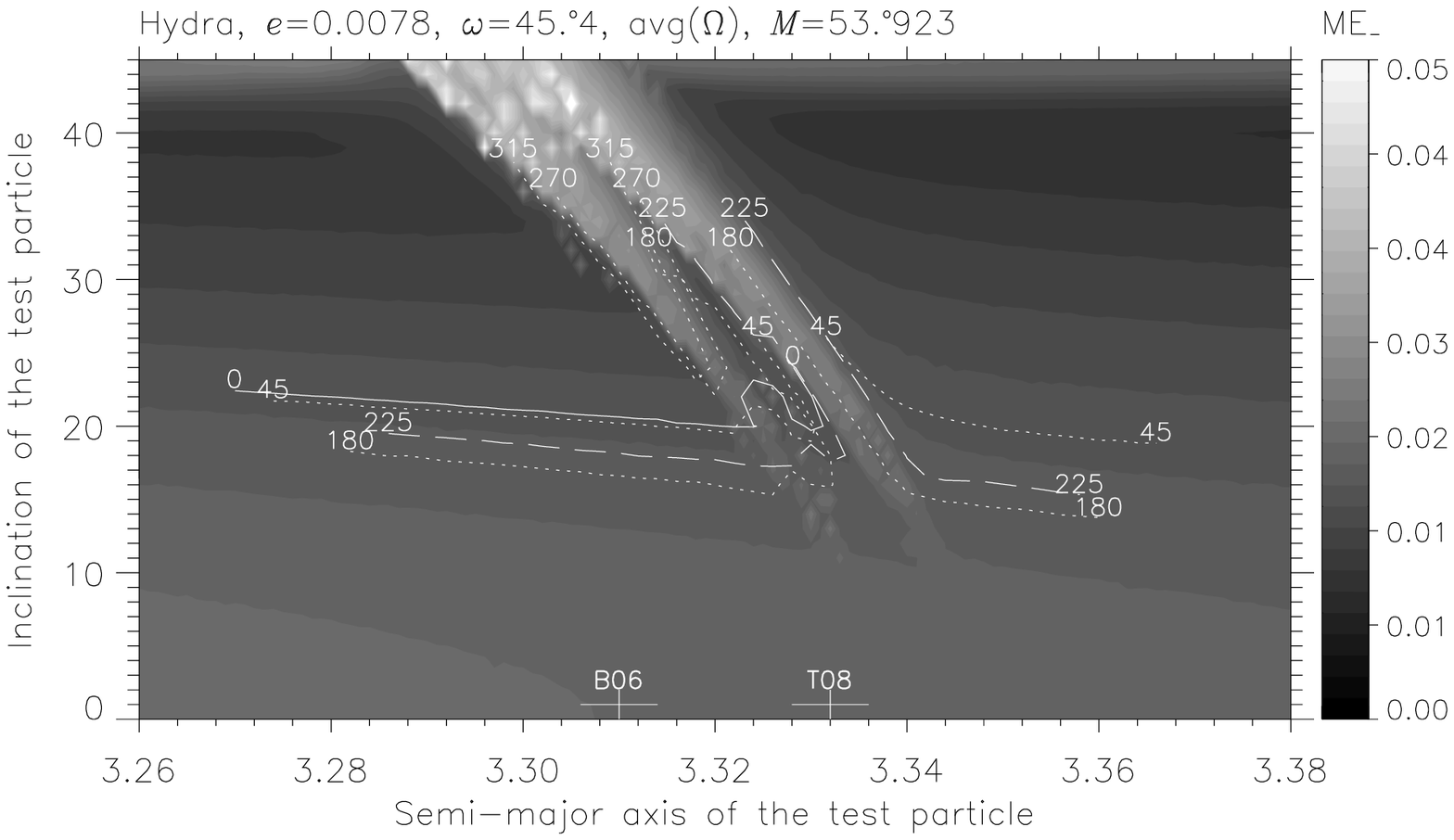}
\caption{The $(a-i)$ stability map for 8 $\Omega$ values for Nix (upper) and Hydra (lower). In the upper panel along the contour curves $\mathrm{ME}_-=0.08$, while for Hydra $\mathrm{ME}_-=0.017$.}
\label{figure7}
\end{figure*}

In Fig. \ref{figure7} the maximum stability maps for $\Omega$ for both moons are presented: the upper panel for Nix, the lower one for Hydra. The $\Omega=135^{\circ}$ map shows that T08 of Nix is in the "sting" of the 4:1 resonance while all others are at larger semimajor axes (Fig. \ref{figure7} upper panel). In the case of Hydra there is no sign of any active resonance in the vicinity of T08, all the observable effects are above 10 degrees. The most prominent features belong to $\Omega=180^{\circ}\,225^{\circ}\,45^{\circ}$ and to $0^{\circ}$.
\par
We have performed runs for $e=0$ in order to estimate the effect of the small eccentricities. It turned out that no observable difference could be seen when comparing it with the eccentric case.
\par
The maximum stability map for Nix with initial $e=0.1$ predicts a very narrow stable zone between 0 and $i \le 2^{\circ}$ at the position of T08. According to the upper panel of Fig. \ref{figure4} the motion with initial $e > 0.15$ is completely unstable.
\par
The 6:1 resonance with initial $e=0.1$ is stronger and the T08 position of Hydra is very close to it. The investigated domain is stable up to $45^{\circ}$, the $\mathrm{ME}_-$ values rise above 0.1 beyond $40^{\circ}$ but no escape occurs. Interestingly the parameter space is more stable for higher eccentricity ($e=0.2$) and unstable motion occurs only for $e \ge 0.3$, although in accordance with the lower panel of Fig. \ref{figure4} orbits with \mbox{$a > 3.28$ A} are stable if the inclination is small $(i \le 10^{\circ})$. In the case of $e=0.4$ shear instability characterize the system.

\section{Conclusions}

The phase space around the two small moons, Nix and Hydra of the Pluto--Charon system was studied in detail using the framework of the elliptic restricted three-body problem. As initial conditions the orbital elements of T08 were used and were varied to map the moons' phase space. We studied the effect of increasing the mass parameter and found out that 25\% change only marginally influences the results: the most significant effect is a shift of the structures along the horizontal axis. This might be important in the cases of resonances.
\par
The former work of \cite{Nagy2006} used the circular problem, but present observations favor elliptic orbit of Charon (T08). To compare the elliptic and the circular case computations were performed in both models. Several important differences were observed: ({\it i}) the unstable zone is much larger in the elliptic case, ({\it ii}) the center of the 4:1 resonance is shifted from $a \approx 2.544$ A to $a \approx 2.564$ A, although its shape did not change significantly, ({\it iii}) the 6:1 resonance is shifted too and it has a completely different shape and ({\it iv}) the resonances become stronger for low eccentricities in the elliptic case.
\par
The present positions of the moons (denoted by T08 on the figures) are in stable regions both on the $(a-e)$ and $(a-i)$ parameter spaces. On the stability maps the structures related to the 4:1 and the 6:1 mean motion resonances are clearly visible but none of them contains any of the moons. These results are in line with those of B06 and T08.
\par
"Maximum" $(a-e)$ stability maps were created for both moons. The map for Nix shows the possibility of active resonance in the case of $\omega=157.^{\circ}9$ or $\Omega=135^{\circ}$. For Hydra the "maximum" stability map did not show this possibility, since the 6:1 resonance is very weak in the vicinity of the moon. It was also showd that the planar and spatial cases are almost identical, when the inclinations are very small, like those of the moons.
\par
Analogous maps were presented for the $(a-i)$ plane which confirmed our previous findings. Interestingly in the case of Nix the contour line belongs to 0.08, which is the double of that belonging to the $(a-e)$ maps.

\section*{Acknowledgments}

This work was supported by the Hungarian Scientific Research Fund, grant no. OTKA A017/09. Most of the numerical integrations were accomplished on the NIIDP (National Information Infrastructure Development Program) cluster grid system in Hungary.

\label{lastpage}

\end{document}